\documentstyle[12pt]{article}

\title{{\bf Polarization of the nuclear medium and RPA-type calculations
in $K^+$ scattering from nuclei}}
\author{ J.C. CAILLON and J. LABARSOUQUE\\
Laboratoire de Physique Th\'eorique$^{\dag}$,
 Universit\'e Bordeaux I\\ rue du
Solarium, 33175 Gradignan Cedex, France}
\date{}
\begin{document}
\begin{titlepage}
\maketitle
\thispagestyle{empty}
\begin{abstract}
In the calculation of the $K^+$-nucleus cross sections, the coupling
of the mesons exchanged between the $K^+$ and the target nucleons
to the polarization of the Fermi sea has been
taken into account. This polarization has been calculated in the
one-loop approximation but summed up to all orders (RPA-type calculation).
This effect is found to be rather important but does not improve the
agreement with experiment.
\end{abstract}

LPTB-93-9  \\
E-mail LABARS@FRCPN11.IN2P3.FR \\
\footnotesize{$\dag$ Unit\'e associ\'ee au CNRS 764}
\end{titlepage}
 \newpage
 \section{introduction}
 It is well known now for a long time that the $K^+$ mesons, which are the
 weakest of the strong-interacting probes, penetrate deeply into the high
 density regions of the nucleus\cite{do,clr,al}. In these regions,
 dynamical objects like
 the nucleons and mesons would behave differently than in vacuum.\\
 It has been
 shown\cite{br,cl} that the $K^+$-nucleus cross-sections would be
 particularly sensitive to the in-medium properties of the $\sigma$, $\omega$
 and $\rho$ mesons whose exchange provides the dominant part of the
 $K^+$-nucleon interaction. This sensitivity appears considering
 that, in the medium, the mesons behave like free ones but with
 density-dependent effective masses. For a better understanding of this
 effect, it would be interesting to go beyond
 such an approximation.
 In the framework of a nuclear matter
 built up of baryons and mesons, it seems reasonable to consider that, to
 a large extent, this dressing of the mesons can be interpreted as a coupling
 with excitations of the Fermi sea.                       \\
\indent In this paper we have performed a calculation of the $K^+$-nucleus
 cross sections taking into account the coupling of the $\sigma$, $\omega$ and
 $\rho$ mesons exchanged between the $K^+$ and the target nucleons,
 to the polarization of the Fermi sea. This polarization has been
 calculated in the one-loop approximation but summed up to all orders
 in the mesons propagators (RPA-type calculation, see fig 1). \\
\indent We have analyzed these effects on the ratio $R_T$ of $K^+-^{12}C$ to
$K^+-d$ total cross sections which has been measured\cite{e,x,p}
from $400 MeV/c$ to $900 MeV/c$.\\
\begin{equation}
\label{r}
R_T = \frac{\sigma_{tot}(K^+-^{12}C)}{6 \cdot \sigma_{tot}(K^+-d)}
\end{equation}
As emphasized by many authors, this ratio is less
sensitive to experimental and theoretical uncertainties than, for example,
differential cross sections, and thus more transparent to the underlying
physics. \\
In section 2 we present the formalism used to calculate the $K^+$-nucleus
cross sections and in section 3 how we have modified the kernel of the KN
interaction
to take into account the polarization of the medium. The results are then
discussed in section 4.

\section{The $K^+$-nucleus cross-sections}
The optical potential describing the elastic scattering of $K^+$ mesons from
nuclei has been constructed by folding the nuclear proton and neutron
densities, $\rho_p$ and $\rho_n$, with the density-dependent $K^+$-nucleon
amplitude assumed to be the same as in infinite nuclear matter at the
same density.
 The Fermi averaging, which contributes very
weakly
to the total cross section, has been neglected, but the nonlocalities and
off-shell
behaviour of the $KN$ boson exchange model have been preserved. This leads
in momentum space to:
\begin{eqnarray}
\label{u}
\langle \overrightarrow{k'} | U | \overrightarrow{k} \rangle & = & \int
e^{i(\overrightarrow{k'} - \overrightarrow{k})\cdot \overrightarrow{r}}
[Z \langle \overrightarrow{k'} | t_{Kp}(\rho_p(\vec{r})) | \overrightarrow{k}
\rangle \rho_p(\vec{r}) \nonumber \\
 & & \mbox{}+N \langle \overrightarrow{k'} | t_{Kn}(\rho_n(\vec{r})) |
\overrightarrow{k} \rangle \rho_n(\vec{r})] d^3r
\end{eqnarray}
\indent The point-like proton distribution of the $^{12}C$ nucleus required in
the
present analysis is that deduced, after the proton finite-size correction has
been made, from the electron-scattering charge density of Sick and McCarthy
\cite{20} and we have chosen equal n and p-distributions. For the deuteron,
the optical potential has been calculated in the same way and we have used
densities deduced from the Hulth\'en wave function\cite{21}.\\
\indent The scattering amplitude is then calculated
 by solving a relativistic Lippman-Schwinger type equation in momentum
space
\begin{equation}
\label{t}
\langle \overrightarrow{k'} | T(E) | \overrightarrow{k} \rangle  =
\langle \overrightarrow{k'} | U(E) | \overrightarrow{k} \rangle + \int
\frac{\langle \overrightarrow{k'} | U(E) | \overrightarrow{k"} \rangle
\langle \overrightarrow{k"} | T(E) | \overrightarrow{k} \rangle }
{E - E_A(\overrightarrow{k"}) - E_K(\overrightarrow{k"}) + i\epsilon} d^3k"
\end{equation}
where $E_A(\overrightarrow{k}) = \sqrt{\overrightarrow{k}^2 + M_A^2}$ and
$E_K(\overrightarrow{k}) = \sqrt{\overrightarrow{k}^2 + m_K^2}$ are the nucleus
and $K^+$ energies. This equation is solved by partial wave decomposition,
discretization and matrix inversion.\\
The $K^+$-nucleus total cross section is deduced from the forward
$K^+$-nucleus elastic scattering amplitude using the optical theorem.\\
For the $K^+$-nucleon amplitude in free space, we have used here the full
Bonn boson exchange model\cite{bo} which is actually one of the more elaborate
descriptions of the KN interaction.
 This $K^+N$ interaction
includes single particle exchange ($\omega, \rho, \sigma, \Lambda, \Sigma,
Y^*$) but also fourth-order diagrams involving $\pi$ and $\rho$ exchange with
N, $\Delta$, K and K$^*$ intermediate states. The solution B1, which provides
the best agreement with experimental data, will be used here. We think that
this agreement, though not perfect, is good enough to expect that the main part
of the physics has been taken into account.
 In nuclear matter, this amplitude has been modified in order to
take into account the coupling of the $\sigma$, $\omega$ and $\rho$ mesons,
whose exchange provides the dominant part of the KN interaction, to the
polarization of the medium. The heavier particles exchanged lead to very
short-ranged processes less influenced by the nuclear environment.

\section{The in-medium KN amplitude}
The meson propagators taking into account the polarization of the medium
can be written\cite{cps} (the Feynman propagators contain in addition
a factor $i$ which has been dropped here for simplicity):
\begin{equation}
\label{g1}
G_{\sigma \sigma} = \frac{q^2 - m_{\omega}^2 - \Pi_{q}^{\omega}
+  \Pi_{\eta}^{\omega}}{(q^2 - m_{\sigma}^2 -  \Pi_{\sigma \sigma})
(q^2 - m_{\omega}^2 - \Pi_{q}^{\omega} +  \Pi_{\eta}^{\omega})
+  \Pi_{\sigma \omega}^2}
\end{equation}
\begin{equation}
\label{g2}
G_{\omega \omega}^{\mu \nu} = \frac{-g^{\mu \nu} + \eta^{\mu}\eta^{\nu}}
{q^2 - m_{\omega}^2 - \Pi_{q}^{\omega}} - \eta^{\mu}\eta^{\nu} \frac{
 q^2 - m_{\sigma}^2 - \Pi_{\sigma \sigma}}
 {(q^2 - m_{\sigma}^2 -  \Pi_{\sigma \sigma})
(q^2 - m_{\omega}^2 - \Pi_{q}^{\omega} +  \Pi_{\eta}^{\omega})
+  \Pi_{\sigma \omega}^2}
\end{equation}
\begin{equation}
\label{g3}
G_{\sigma \omega}^{\mu} = \frac{- \eta^{\mu} \Pi_{\sigma \omega}}
 {(q^2 - m_{\sigma}^2 -  \Pi_{\sigma \sigma})
(q^2 - m_{\omega}^2 - \Pi_{q}^{\omega} +  \Pi_{\eta}^{\omega})
+  \Pi_{\sigma \omega}^2}
\end{equation}
\begin{equation}
\label{g4}
G_{\rho \rho}^{\mu \nu} = (-g^{\mu \nu} +\frac{ \eta^{\mu}\eta^{\nu}}{\eta^2})
\frac{1}{q^2 - m_{\rho}^2 - \Pi_{q}^{\rho}} - \frac{\eta^{\mu}\eta^{\nu}}
{\eta^2} \frac{1}{q^2 - m_{\rho}^2 - \Pi_{q}^{\rho} +  \Pi_{\eta}^{\rho}}
\end{equation}
where $\Pi_{\sigma \sigma}$ is the polarization in the $\sigma$ channel and
the polarization in the $\omega$, $\sigma \omega$ and $\rho$ channels have been
decomposed as:
\begin{equation}
 \Pi_{\omega \omega}^{\mu \nu} = -(g^{\mu \nu} - \frac{q^{\mu}q^{\nu}}
 {q^2}) \Pi_{q}^{\omega} + \frac{{\hat{\eta}}^{\mu}{\hat{\eta}}^{\nu}}
 {{\hat{\eta}}^2} \Pi_{\eta}^{\omega}
 \end{equation}
 \begin{equation}
 \Pi_{\sigma \omega}^{\mu} = \frac{{\hat{\eta}}^{\mu}}
 {({\hat{\eta}}^2)^{1/2}} \Pi_{\sigma \omega}
 \end{equation}
\begin{equation}
 \Pi_{\rho \rho}^{\mu \nu} = -(g^{\mu \nu} - \frac{q^{\mu}q^{\nu}}
 {q^2}) \Pi_{q}^{\rho} + \frac{{\hat{\eta}}^{\mu}{\hat{\eta}}^{\nu}}
 {{\hat{\eta}}^2} \Pi_{\eta}^{\rho}
 \end{equation}
 Here $\eta^{\mu}$ which describes the uniform motion of the medium, is such
 that, in the nuclear matter rest frame we have $\eta^{\mu} = (1,0,0,0)$ and
 we have noted
 \begin{equation}
 {\hat{\eta}}^{\mu} = {\eta}^{\mu} - \frac{(q\cdot\eta) q^{\mu}}{q^2}
 \end{equation}
 For simplicity, the isospin indices have been omitted in the case of the
 $\rho$-meson.
 In the medium, the $\sigma$ and the longitudinal part of the $\omega$
 are strongly mixed and, consequently, the structure of the
 propagators (eq \ref{g1},\ref{g2},\ref{g3}) is
 rather complicated. However, these propagators can be written in a form
 which makes the physics more apparent.
 \begin{equation}
 \label{gss}
G_{\sigma \sigma} = \frac{m_+^2 - m_{\omega_L}^2 }{m_+^2 - m_-^2 }
 \frac{1}{q^2 - m_+^2}
+ \frac{m_-^2 - m_{\omega_L}^2 }{m_-^2 - m_+^2 } \frac{1}{q^2 - m_-^2}
\end{equation}
\begin{equation}
\label{goo}
G_{\omega \omega}^{\mu \nu} = \frac{-g^{\mu \nu} + \eta^{\mu}\eta^{\nu}}
{q^2 - m_{\omega_T}^2 } - \eta^{\mu}\eta^{\nu}
(\frac{m_+^2 - m_{\sigma^*}^2 }{m_+^2 - m_-^2 } \frac{1}{q^2 - m_+^2}
+ \frac{m_-^2 - m_{\sigma^*}^2 }{m_-^2 - m_+^2 } \frac{1}{q^2 - m_-^2})
\end{equation}
\begin{equation}
\label{gso}
G_{\sigma \omega}^{\mu} = \frac{- \eta^{\mu} \Pi_{\sigma \omega}}
{m_+^2 - m_-^2 }( \frac{1}{q^2 - m_+^2} -  \frac{1}{q^2 - m_-^2})
 \end{equation}
 where $m_{\sigma^*}$ is the $\sigma$-mass modified by the polarization
 in the $\sigma$ channel
\begin{equation}
 m_{\sigma^*}^2 = m_{\sigma}^2 + \Pi_{\sigma \sigma}
 \end{equation}
  $m_{\omega_L}$ and $ m_{\omega_T}$ are the $\omega$-mass modified
  respectively by the longitudinal and transverse part of the polarization
in the $\omega$ channel
\begin{equation}
 m_{\omega_L}^2 =  m_{\omega}^2 + \Pi_q^{\omega} - \Pi_{\eta}^{\omega}
 \end{equation}
\begin{equation}
 m_{\omega_T}^2 =  m_{\omega}^2 + \Pi_q^{\omega}
 \end{equation}
and $m_+$, $m_-$ are defined by
\begin{equation}
m_{\pm}^2 = \frac{1}{2} ( m_{\sigma^*}^2 + m_{\omega_L}^2 \pm
[ (m_{\omega_L}^2 - m_{\sigma^*}^2)^2 - 4 \Pi_{\sigma \omega}^2]^{1/2} )
 \end{equation}
Thus, we can see that, apart the transverse part of the $\omega$ not affected
by the $\sigma -\omega$ mixing,
the propagators appearing in the medium are those of two particles of mass
$m_+$ and $m_-$. These two particles, arising from the mixing of the $\sigma$
and of the longitudinal part of the $\omega$, can couple either as a scalar
or as a vector, giving rise to the three propagators $G_{\sigma \sigma}$,
$G_{\omega \omega}^{\mu \nu}$ and $G_{\sigma \omega}^{\mu}$.\\
The propagator of the $\rho$-meson, more simple, can be written directly
\begin{equation}
\label{grr}
G_{\rho \rho}^{\mu \nu} = (-g^{\mu \nu} + \frac{ \eta^{\mu}\eta^{\nu}}{\eta^2})
\frac{1}{q^2 - m_{\rho_T}^2 } - \frac{\eta^{\mu}\eta^{\nu}}
{\eta^2} \frac{1}{q^2 - m_{\rho_L}^2}
\end{equation}
where $m_{\rho_L}$ and $m_{\rho_T}$ are defined as in the case of the
$\omega$-meson
\begin{equation}
 m_{\rho_L}^2 =  m_{\rho}^2 + \Pi_q^{\rho} - \Pi_{\eta}^{\rho}
 \end{equation}
\begin{equation}
 m_{\rho_T}^2 =  m_{\rho}^2 + \Pi_q^{\rho}
 \end{equation}
The expressions for the polarizations $\Pi_{\sigma \sigma}$,
 $\Pi_{q}^{\omega}$, $\Pi_{\eta}^{\omega}$, $\Pi_{\sigma \omega}$,
 $\Pi_q^{\rho}$ and $\Pi_{\eta}^{\rho}$
 at the one-loop approximation can be found in ref\cite{cps}.\\
 \indent The KN amplitude used in this work (the Bonn boson exchange
 model), built using the Time Dependent Perturbation Theory, is fully
relativistic but not in a covariant form. Thus,
 to be able to take into account the polarization effects described
 above in the Feynman covariant theory, we have to establish a
 correspondance procedure between the two formalisms. We have used here
 the procedure already employed by Kotthoff et al\cite{ko} in the NN case,
 i.e. we have required that the KN meson-exchange potential in the
 Feynman formalism $V^F$ be related to that in the Time Dependent
 Perturbation Theory $V^{TD}$ by
 \begin{equation}
 V^F(E',\vec{p'};E,\vec{p}) = \frac{1}{2}(V^{TD}(z = E + \omega_K)
 + V^{TD}(z = E' + \omega_K'))
 \end{equation}
 where $E = \sqrt{M_N^2 + \vec{p}^2}$, $E' = \sqrt{M_N^2 + \vec{p'}^2}$,
 $\omega_K = \sqrt{m_K^2 + \vec{p}^2}$ and $\omega_K' =
 \sqrt{m_K^2 + \vec{p'}^2}$.
 This can be realized considering the expressions obtained for
 the Feynman propagator (eq \ref{gss},\ref{goo},\ref{gso},\ref{grr})
 which can be written
 \begin{equation}
 G = \sum_{i} \frac{\alpha_i}{q^2 - m_i^2 }
 \end{equation}
 and then performing an off-energy-shell extrapolation\cite{ko}
 of each term in order to obtain
 the corresponding two different time-ordered processes
 \begin{equation}
 G \longrightarrow G_{TD} = \sum_{i} \frac{\alpha_i}{2 \omega_i}
 (\frac{1}{z - E - \omega_K' - \omega_i} + \frac{1}{z - E' - \omega_K
 - \omega_i})
 \end{equation}
 with $ \omega_i^2 = \vec{q}^2 + m_i^2 $.
\section{Results and discussion}
Considering that the dominant contribution to the forward $K^+$-nucleus
scattering
amplitude and thus to the $K^+$-nucleus total cross section comes from
forward $K^+$-nucleon scattering, we have estimated that,
in such a region of small energy-momentum
transfer, the calculation of the polarization
at the one-loop order (see fig 1,a) should be a reasonable approximation.
Therefore, we have used here the
expressions obtained in that way by Celenza, Pantziris and Shakin\cite{cps}.
However, in order to take into account, in addition, the main effect of the
nuclear field on the nucleons Dirac
spinors, we have replaced in the polarization
the nucleon mass M by an effective mass M$^*$. The density dependence of
this effective nucleon mass has been chosen in such a way we obtain at
saturation a mass decreased by 15\%, which is the value often considered
as the most realistic\cite{ms}. More precisely, we have taken
\begin{equation}
 M^*(\rho) = M (1 - 0.15\frac{\rho}{\rho_0})
\end{equation}
and we have verified that smooth variations from this linear dependence
don't change significantly the results.\\
For consistency, the coupling constants and form-factors of the
B1 Bonn interaction have been used at each vertex.\\
The $\sigma$ and $\omega$ exchange terms in the KN interaction cannot be
longer considered independently and have to be replaced by a
``$\sigma$-$\omega$''exchange which has a scalar, a vector, a scalar-vector
and a vector-scalar part according to the character of its coupling to the
K$^+$ and to the nucleon.
The  masses describing the in-medium propagation of this ``$\sigma$-$\omega$''
meson are now $m_{\omega_T}$,
$m_+$ and $m_-$. The $\omega$-mass in
the transverse channel, $m_{\omega_T}$, is weakly modified in nuclear
matter (a few percents), and the main effect comes from the $\sigma$ and
$\omega$-longitudinal exchanges which are strongly coupled. As discussed in the
preceding section, the propagators appearing now in this sector are not
those of the
$\sigma$ and $\omega$ mesons, but propagators of particles with mass $m_+$
and $m_-$ different from $m_{\sigma}$ and $m_{\omega}$.
For example, in the forward direction which gives the dominant
contribution here,
 these masses are such that their real part (fig 2) are approximately
 7\% lower than
 the $\sigma$-mass for $m_-$ and 10\% higher than the $\omega$-mass for $m_+$,
 for densities higher than 20\% of the saturation density.
The imaginary parts grow from zero in vacuum to $\sim$200 MeV at saturation.
Moreover, the weights of the propagators of these ``particles'' with mass
$m_+$ and $m_-$ entering in the scalar, vector, scalar-vector and
vector-scalar parts of the ``$\sigma-\omega$'' exchange are varying
with density.
Thus, in the medium, the KN scattering amplitude will be appreciably different
than in free space.\\
In symmetric nuclear matter, the $\rho$-meson exchange contributes weakly
to the KN interaction and its coupling to the polarization of the medium
modifies very weakly the $K^+$-nucleus forward amplitude. However, this
coupling has been included here for completeness.\\
The $K^+$-$^{12}$C total cross section has been calculated using this
medium-modified KN amplitude. For the calculation of the
$K^+$-deuteron cross section, since the densities involved are small and since
in a nucleus made up of two nucleons it is not possible to excite more than
one particle-hole pair, we have used the free-space KN interaction.\\
The ratio $R_T$ obtained when the polarization of the nuclear medium is
taken into account as indicated above, is shown fig.3, curve b. We can
see that this effect is important since the $R_T$ ratio is now, in average,
$\sim$5\%
lower than that calculated using the free-space KN interaction ($\sim$10\%
at 400 MeV/c and $\sim$2\% at 900 MeV/c).    \\
Unfortunately, the agreement between theory and experiment is not
improved. However, we should claim that medium effects as elementary as
particle-hole
excitations, which are known to be important in the whole range of nuclear
physics and which have been found to be important here, cannot be ignored in
 this
problem, even if the solution has to be searched elsewhere.
   
\newpage

\begin{center}
{\bf Figure captions}
\end{center}
\bigskip

Fig 1:a) Polarization of nuclear matter at the one-loop approximation
in the $\sigma$-$\sigma$, $\sigma$-$\omega$, $\omega$-$\sigma$ and
$\omega$-$\omega$ channels; b)
RPA-sum contributing to the
 kernel of the KN interaction in nuclear
matter arising from $\sigma$ and $\omega$ exchange; the dashed lines
represent $\sigma$ or $\omega$ mesons indifferently.\\

Fig 2: Real part of the masses $m_+$ and $m_-$ arising from the mixing
of the $\sigma$ and of the longitudinal part of the $\omega$ in nuclear
matter calculated in the forward direction ($\nu$ = 0, $\|\vec{q}\|$ =
 10 MeV/c)\\

Fig.3: Ratio $R_T$ of the $K^+-^{12}C$ and $K^+-d$ total cross sections
as a function of $p_{lab}$ calculated, curve (a): with the free-space $K^+N$
interaction, curve (b): with a density-dependent $K^+N$ interaction taking
into account the coupling of the $\sigma$, $\omega$ and $\rho$ mesons with
the polarization of nuclear matter calculated at the one-loop approximation.
  The experimental
points are taken from ref.\cite{x} (circles), from ref.\cite{e} (squares) and
from ref.\cite{p} (triangles).
\end{document}